\documentclass[11pt]{article}
\usepackage{setspace}
\usepackage{color}
\usepackage{amsmath,cite}
\usepackage{amsfonts}
\usepackage{verbatim}
\usepackage{amssymb}
\usepackage{graphicx,bm}
\usepackage{graphicx,cite,enumerate}
\usepackage{authblk} 
\usepackage{bm}

\definecolor{darkgreen}{rgb}{0,0.35,0}

\providecommand{\U}[1]{\protect\rule{.1in}{.1in}}
\newcommand{\be}{\begin{equation}}
\newcommand{\ee}{\end{equation}}
\newcommand{\beqn}{\begin{eqnarray}}
\newcommand{\eeqn}{\end{eqnarray}}
\newcommand{\pa}{\partial}

\newcommand{\ottG}{\kappa}
\newcommand{\sedG}{2\kappa}
\newcommand{\quattG}{\frac{\kappa K}{2}}
\def \e {\mbox{\boldmath{$e$}}}
\def \x {\mbox{\boldmath{$\xi$}}}

\begin{document}

\title{Extensions of the generalized hedgehog ansatz for the Einstein--nonlinear $\sigma$-model system: black holes with NUT, black strings and time-dependent solutions}

\author[1]{Alex Giacomini\thanks{alexgiacomini@uach.cl}}
\author[2]{Marcello Ortaggio\thanks{ortaggio(at)math(dot)cas(dot)cz}}

\affil[1]{Instituto de Ciencias F\'{\i}sicas y Matem\'aticas, Universidad Austral de Chile, Edificio Emilio Pugin, cuarto piso, Campus Isla Teja, Valdivia, Chile}
\affil[2]{Institute of Mathematics of the Czech Academy of Sciences, \newline \v Zitn\' a 25, 115 67 Prague 1, Czech Republic}

\maketitle

\begin{abstract}

We consider a class of ans\"atze for the construction of exact solutions of the
Einstein-nonlinear $\sigma$-model system with an arbitrary cosmological constant in (3+1)
dimensions. Exploiting a geometric interplay between the $SU(2)$ field and Killing vectors of the spacetime reduces the matter field equations to a single scalar equation (identically satisfied in some cases) and simultaneously simplifies Einstein's equations. This is then exemplified over various classes of spacetimes, which allows us to construct stationary black holes with a NUT parameter and uniform black strings, as well as time-dependent solutions such as Robinson-Trautman and Kundt spacetimes, Vaidya-type radiating black holes and certain Bianchi~IX cosmologies. In addition to new solutions, some previously known ones are rederived in a more systematic way.

\end{abstract}


\section{Introduction}

Nonlinear $\sigma$-models have many important applications, e.g., in quantum field theory and statistical mechanics, cf., e.g., \cite{ManStu_book,Nair_book} and references therein.  In particular, the $SU(2)$ nonlinear $\sigma$-model is a low energy effective model for pion interactions. This model can be extended by adding a so called Skyrme term in order to admit stable solitonic solutions. In this way the Skyrme model can describe pions as well as baryons in (3+1) dimensions \cite{ManStu_book,Nair_book}. The field equations can be described as a system of non linear partial differential equations which are extremely difficult to solve. Not surprisingly, the coupling to gravity makes the problem even more complicated and additionally requires also Einstein's equations to be solved. For this reason, a lot of work has been devoted to numerical studies, cf., e.g., \cite{LucMos86,num1,num2,num3,num4,num5,num6,num7,num8,num9,num10,num11,num12} and references therein.

Finding exact solutions of the Einstein-nonlinear $\sigma$-model or Einstein-Skyrme system is thus not an easy task. Nevertheless, having in mind a specific physical system, the choice of a good ansatz for the matter field and the spacetime metric may reduce the complexity of the problem. A well-known example is given by the hedgehog ansatz in the presence of spherical symmetry, considered in this context (without backreaction) in \cite{LucMos86,ManRub86} and further used (also with backreaction) in a number of works, see, e.g., \cite{VolGal99,ShiSaw05,Gussmann17} for reviews and more references. A generalizations of the hedgehog ansatz for the Skyrme-Einstein system beyond spherical symmetry has been constructed in \cite{CanMae13} and applied to spacetimes which possess plane or hyperbolic symmetry or are stationary and axisymmetric (see \cite{CanSal13} for earlier results for the $SU(2)$ nonlinear $\sigma$-model). With similar techniques, topologically non-trivial solutions in curved spacetimes have been obtained in \cite{AyoCanZan16}.

In the present paper we construct further extensions of the ansatz used in \cite{CanSal13,CanMae13,AyoCanZan16}.\footnote{The ansatz of \cite{CanSal13,CanMae13,AyoCanZan16} is called a ``generalization'' of the hedgehog ansatz in that it applies to certain spacetimes which do {\em not} possess spherical symmetry, as opposed to the standard hedgehog ansatz. In turn, the various ans\"atze considered in the present paper can all be generically referred to as ``extensions of the generalized hedgehog ansatz'' in the sense that they are inspired by \cite{CanSal13,CanMae13,AyoCanZan16} and take advantage of similar (but not identical) interplays between the internal space and spacetime coordinates (and can also apply to spacetimes with less or no symmetries, at least in some cases, cf. sections~\ref{subsec_RT_flat}, \ref{subsec_K_flat}, \ref{subsec_RT_spher}, \ref{subsec_K_spher}).} For simplicity, we restrict ourselves to the $SU(2)$ nonlinear $\sigma$-model, without the Skyrme term.\footnote{It is worth pointing out that one possible way to evade Derrick's scaling argument (which prohibits static solitonic solutions for the nonlinear $\sigma$-model) \cite{ManStu_book} is to couple it with gravity. For this reason, the self-gravitating $SU(2)$ nonlinear $\sigma$-model without Skyrme term has its own physical interest.} On the one hand, we will show that certain solutions obtained in various earlier papers can be in derived in a unified, more sistematic way. On the other hand, we will demonstrate that the same method can also be used to construct some new solutions, and further allows one to drop the requirement of spacetime symmetries assumed in previous works (at least under certain circumstances, as we shall explain in the following). This enables one, in particular, to construct time-dependent Robinson-Trautman or Kundt spacetimes \cite{Stephanibook} sourced by the $SU(2)$ field. In passing, we will additionally observe that some of the proposed ans\"atze correspond to truncations of the theory which make it effectively equivalent to Einstein gravity coupled to (depending on the concrete ansatz) two axionic fields, a single scalar field or an anisotropic fluid. Some of the obtained solutions can thus be reinterpreted also in those contexts, and viceversa.

The structure of the paper is as follows. In section~\ref{sec_action} we briefly review the theory of interest and discuss two possible parametrizations of the $SU(2)$ fields. Based on those, in section $3$ we present the ans\"atze for the $SU(2)$ fields and for the spacetime geometries that we are going to consider. In sections~\ref{sec_1ai}--\ref{sec_hypers} we classify the exact solutions according to the considered  ans\"atze and present several explicit examples. The concluding section~\ref{sec_concl} is devoted to a brief summary of the results and further comments. An appendix contains a few remarks about test fields in the Kerr spacetime.

\section{The action}

\label{sec_action}

We consider the Einstein-nonlinear $\sigma$-model system in four dimensions in the presence of an arbitrary cosmological constant $\Lambda$. The degrees of freedom of the nonlinear $\sigma$-model are encoded in an $SU(2)$ group-valued scalar field $U$. It arises as a special case of the Einstein-Skyrme theory by setting the Skyrme term to zero \cite{ManStu_book,Nair_book}. The system
action is the sum of a gravitational action and a nonlinear $\sigma$-model action
\begin{equation}
S=S_{\mathrm{G}}+S_{\sigma},  \label{actiontotal}
\end{equation}
where
\be
S_{\mathrm{G}}= \frac{1}{\sedG }\int d^{4}x\sqrt{-g}(\mathcal{R}-2\Lambda
),  \qquad S_{\sigma}= \frac{K}{4}\int d^{4}x\sqrt{-g}\mathrm{Tr}\left( R^{\mu
}R_{\mu }\right) ,  \label{sky2}
\ee
with the Maurer-Cartan 1-form given by
\begin{equation}
R_{\mu }=U^{-1}\nabla _{\mu }U\ , \qquad U\in SU(2) .
 \label{defR}
\end{equation}
Here $\kappa=8\pi G$, $G$ is Newton's constant, the parameter $K$ (of dimension $1/G$) is positive and $\mathcal{R}$ is the Ricci scalar. In our conventions $c=\hbar =1$, the spacetime signature is $(-,+,+,+)$ and Greek indices run over spacetime.

The Einstein equations derived from \eqref{actiontotal} are
\begin{equation}
G_{\mu \nu }+\Lambda g_{\mu \nu }=\ottG T_{\mu \nu },  \label{einstein}
\end{equation}
with
\begin{equation}
 T_{\mu \nu }=-\frac{K}{2}\mathrm{Tr}\left( R_{\mu }R_{\nu }-\frac{1}{2}
g_{\mu \nu }R^{\alpha }R_{\alpha }\right) ,   \label{timunu1}
\end{equation}
while the matter field equations read
\begin{equation}
\nabla ^{\mu }R_{\mu }=0.  \label{nonlinearsigma1}
\end{equation}
It is a standard result (cf., e.g., appendix~E of \cite{Waldbook}) that the latter imply $T^{\mu \nu}_{\phantom{\mu\nu};\nu}=0$.

\subsection{Parametrizations of the $SU(2)$ field}

It is useful to define the three invariant 1-forms $R_{\mu }^{j}$ by writing $R_{\mu }$ as
\begin{equation}
R_{\mu }=iR_{\mu }^{j}\sigma_{j} \qquad (j=1,2,3) ,  \label{Ri}
\end{equation}
where $\sigma _{j}$ are the Pauli matrices. We adopt the
standard parametrization of the $SU(2)$-valued scalar $U(x^{\mu })$
\begin{equation}
U^{\pm 1}(x^{\mu })=Y^{0}(x^{\mu })\mathbb{\mathbf{I}}\pm
iY^{j}(x^{\mu })\sigma_{j}\ , \qquad \left( Y^{0}\right)
^{2}+Y^{i}Y_{i}=1\,,  \label{standnorm}
\end{equation}
where $\mathbb{\mathbf{I}}$ is the $2\times 2$ identity, which with \eqref{defR} and \eqref{Ri} gives
\begin{equation}
 R_{\mu }^{i}=\varepsilon ^{ijk}Y_{j}\nabla _{\mu
}Y_{k}+Y^{0}\nabla _{\mu}Y^{i}-Y^{i}\nabla _{\mu }Y^{0} .  \label{Ri_2}
\end{equation}

The second of \eqref{standnorm} means that $Y^{I}=(Y^{0},Y^{i})$ define a round unit 3-sphere in the internal space. Defining its $S^3$ metric
\begin{equation}
G_{ij}=\delta _{ij}+\frac{Y_{i}Y_{j}}{1-Y^{k}Y_{k}}\   \label{intmetric} ,
\end{equation}
the action (\ref{sky2}) takes the form
\begin{equation}
	S_{\sigma}=-K\int d^{4}x\sqrt{-g}\left[ \frac{1}{2}G_{ij}(\nabla
_{\mu }Y^{i})(\nabla ^{\mu }Y^{j})\right] ,
\end{equation}
while the energy-momentum tensor (\ref{timunu1}) becomes
\begin{equation}
T_{\mu \nu }=K\left( \mathcal{S}_{\mu \nu }-\frac{1}{2}g_{\mu \nu }\mathcal{S}\right) , \label{timunu2}
\end{equation}
with
\begin{equation}
\mathcal{S}_{\mu \nu }=\delta _{ij}R_{\mu }^{i}R_{\nu }^{j}=
\nabla_{\mu }Y^{0}\nabla _{\nu }Y^{0}+\nabla_{\mu
}Y^{i}\nabla _{\nu }Y^{i}=G_{ij}(Y)\nabla _{\mu }Y^{i}\nabla _{\nu }Y^{j} , \qquad \mathcal{S}=g^{\mu\nu}\mathcal{S}_{\mu \nu} .
\label{cuadra1}
\end{equation}
This enables one the rewrite the Einstein equations~\eqref{einstein} in the useful form
\begin{equation}
 {\cal R}_{\mu \nu }=\Lambda g_{\mu \nu }+\ottG K\mathcal{S}_{\mu \nu } ,  \label{einstein2}
\end{equation}
where ${\cal R}_{\mu \nu }$ is the Ricci tensor. The trace of~\eqref{einstein2} clearly reads ${\cal R}=4\Lambda+\ottG K\mathcal{S}$.

For the purposes of the present paper, two different parametrizations of $S^3$ turn out to be convenient.

\subsubsection{Coordinates adapted to two commuting Killing vectors (Hopf coordinates)}

\label{subsubsec_HAG}

A useful set of coordinates $(H,A,G)$ in the internal space is defined by (cf., e.g., \cite{Schroedinger56})
\beqn
 & & Y^{0}=\cos H\sin A , \qquad Y^{1}=\sin H\cos G , \nonumber \\
 & & Y^{3}=\cos H\cos A , \qquad Y^{2}=\sin H\sin G , \label{HAG_def}
\eeqn
where $H\in \lbrack 0,\pi /2]$, while $A\in \left[ 0,2\pi k_{1}\right] $, $
G\in \left[ 0,2\pi k_{2}\right] $ are both periodic Killing coordinates of $
S^{3}$, with $k_{1}$ and $k_{2}$ positive integers.\footnote{There is apparently some redundancy in the above periodicities of the Killing angles, since  the standard choice $k_1=k_2=1$ already covers the whole $S^3$ -- however, this becomes physically meaningful in certain applications \cite{Astorinoetal18}.} The tensor $\mathcal{S}_{\mu \nu }$\ \eqref{cuadra1}
thus takes the form
\begin{equation}
\mathcal{S}_{\mu \nu }=(\nabla _{\mu }H)(\nabla _{\nu }H)+\cos ^{2}H(\nabla
_{\mu }A)(\nabla _{\nu }A)+\sin ^{2}H(\nabla _{\mu }G)(\nabla _{\nu }G).
\label{S_H}
\end{equation}

Next, defining the combinations of the Killing coordinates
\begin{equation}
\Phi _{+}=G+A,\qquad \Phi _{-}=G-A,
\end{equation}
enables one to write the field equations~(\ref{nonlinearsigma1}) as
\begin{eqnarray}
&&\square H-\frac{1}{2}\sin (2H)\nabla \Phi _{+}\cdot\nabla\Phi _{-}=0,
\label{D2H} \\
&&\sin (2H)\square \Phi _{+}+2\nabla H\cdot \left[ \nabla \Phi _{-}+\cos
(2H)\nabla \Phi _{+}\right] =0,  \label{D2ga} \\
&&\sin (2H)\square \Phi _{-}+2\nabla H\cdot \left[ \nabla \Phi _{+}+\cos
(2H)\nabla \Phi _{-}\right] =0,  \label{D2ph}
\end{eqnarray}
where the simple identity $\left( g^{\mu \nu }hf_{,\nu
}\right) _{;\mu }=\nabla h\cdot \nabla f+h\square f$ has been used.

\subsubsection{Hypersferical coordinates}

\label{subsubsec_alFG}

Another useful set of coordinates consists of the standard hyperspherical\footnote{Throughout the paper, we will use the adjective ``hyperspherical'' when referring to $S^3$ and ``spherical'' when referring to $S^2$.}  coordinates $(\alpha,F,G)$, defined by
\begin{align}
Y^{0}& =\cos \alpha , \qquad Y^{i}=n^{i}\sin \alpha \ ,  \nonumber \\
n^{1}& =\sin F\cos G\ ,\quad \ n^{2}=\sin F\sin G\ , \quad n^{3}=\cos F\ ,
\label{pions2}
\end{align}
where $\alpha,F\in[0,\pi]$, while $G\in[0,2\pi]$ is a periodic Killing coordinate. Using these, \eqref{cuadra1} takes the form
\begin{equation}
\mathcal{S}_{\mu \nu }=\left( \nabla _{\mu }\alpha \right) \left( \nabla
_{\nu }\alpha \right) +\sin ^{2}\alpha \left[ \left( \nabla _{\mu }F\right)
\left( \nabla _{\nu }F\right) +\sin ^{2}F\left( \nabla _{\mu }G\right)
\left( \nabla _{\nu }G\right) \right] \ ,  \label{smunu}
\end{equation}
while the field equations (\ref{nonlinearsigma1}) can be rewritten as
\beqn
-\square \alpha +\frac{\sin \left( 2\alpha \right) }{2}\left[ \left( \nabla
_{\mu }F\right) \left( \nabla ^{\mu }F\right) +\sin ^{2}F\left( \nabla _{\mu
}G\right) \left( \nabla ^{\mu }G\right) \right] =0\ ,  \label{fequ1} \\
-\sin ^{2}\alpha \square F-\sin \left( 2\alpha \right)\left(
\nabla _{\mu }F\right) \left( \nabla ^{\mu }\alpha \right) +\sin ^{2}\alpha
\frac{\sin \left( 2F\right) }{2}\left( \nabla _{\mu }G\right) \left( \nabla
^{\mu }G\right) =0\ ,  \label{fequ2} \\
\sin ^{2}F\sin ^{2}\alpha \square G+\sin \left( 2\alpha \right) \sin
^{2}F\left( \nabla _{\mu }G\right) \left( \nabla ^{\mu }\alpha \right) +\sin
\left( 2F\right) \sin ^{2}\alpha \left( \nabla _{\mu }G\right) \left( \nabla
^{\mu }F\right) =0\ .  \label{fequ3}
\eeqn

\section{The ans\"atze}

\label{sec_system}

The system of field equations \eqref{einstein}--\eqref{nonlinearsigma1} is very complicated, in general. However, we will show that they simplify considerably if one chooses a spacetime geometry suitably adapted to the $SU(2)$ field (this extends previous works such as \cite{CanSal13,CanMae13,AyoCanZan16,Astorinoetal18}, more comments will be given throughout the paper). To this end, both the coordinate systems of sections~\ref{subsubsec_HAG} and \ref{subsubsec_alFG} will prove useful, leading to physically interesting solutions and a transparent geometric interpretation thereof. We will denote by $(x^1,x^2,x^3,x^4)$ the generic spacetime coordinates (without specifying which one is time, for now).

\subsection{Ans\"atze in coordinates adapted to two commuting Killing vectors}

\label{subsec_ans1}

Let us start from the internal space coordinates $(H,A,G)$ of \eqref{HAG_def}. First, let us assume $x^1=A$, and that $A$ is a Killing coordinate not only of the internal 3-space, but also of the four-dimensional spacetime. This gives $\sqrt{-g}\square A=(\sqrt{-g}g^{A\hat\mu})_{,\hat\mu}$, where $\hat\mu=2,3,4$. If one further assumes that $g^{A\hat\mu}=0$ for all coordinates $x^{\hat\mu}$ (except, possibly, for additional Killing coordinates) then $\square A=0$. Similar considerations apply to $G$, which can be taken as a second Killing coordinate.

To summarize, we have defined two of the spacetime (Killing) coordinates as
\be
 x^1=A , \quad x^2=G ,
\ee
and assumed the metric conditions
\be
 g^{A 3}=g^{A 4}=0=g^{G 3}=g^{G 4} .
\label{ans1_metric}
\ee
This gives rise to various adapted spacetime geometries for which $\square A=0=\square G$ ($\Leftrightarrow\square\Phi _{\pm}=0$) identically. Among these, we will consider the following possible subcases.

\begin{enumerate}[(i)]
	\item\label{H=const} Toroidal ansatz: $H=$const$\in(0,\pi/2)$. This means that the $SU(2)$ field takes a toroidal configuration $S^1\times S^1$ (unless $A$ or $G$ are further restricted) and
	\be
		\mathcal{S}_{\mu \nu }=\cos^2 H (\nabla _{\mu }A)(\nabla _{\nu }A)+\sin^2 H (\nabla _{\mu }G)(\nabla _{\nu }G) .
		\label{S_tor}
 \ee		
	The special case $H=\pi/4$ corresponds to the two $S^1$ having the same radius. We observe that here the two scalars $A$ and $G$ can be effectively reinterpreted as free axionic fields, since the matter term in Einstein's equations~\eqref{einstein2} reduces to the axionic one (cf., e.g., \cite{Caldarellietal17}), while $\square A=0=\square G$ by construction.

	\item\label{H=0} Circular ansatz: $H=0$. This gives
	\be
	 \mathcal{S}_{\mu \nu }=(\nabla _{\mu }A)(\nabla _{\nu }A) ,
	\ee
	and includes, in particular, the ``pionic vacuum'' when $A=\pi/2$. Here $A$ acts as a single axion (or a free, massless dilaton, cf. \cite{CanMae13} for more comments), while $G$ becomes a stealth field (i.e., it does not contribute to $T_{\mu\nu}$) -- $A$ becomes a null field (pure radiation) in the special subcase $g^{AA}=0$.

	\item\label{H_generic} Hypersferical ansatz: $x^4=H$ and, by \eqref{ans1_metric}, $g^{HA}=0=g^{HG}$. Here all the $SU(2)$ fields are dynamical and $\mathcal{S}_{\mu \nu }$ is fully general, i.e., of the form \eqref{S_H}. Among the ans\"atze considered in this paper, this one will be the only one giving rise to a non-zero topological charge (cf. \cite{AyoCanZan16,Canforaetal19} for comments related to the solutions of section~\ref{sec_hypers}).\footnote{In general, $SU(2)$-valued scalar fields may possess a non-trivial topological charge which, mathematically, is a suitable homotopy class or winding number (cf., e.g., \cite{ManStu_book}).}
	
\end{enumerate}

The extra assumptions \eqref{H=const}, \eqref{H=0} or \eqref{H_generic} imply $\nabla H\cdot\nabla\Phi _{\pm}=0$, so that the field equations~\eqref{D2ga} and \eqref{D2ph} become identities. In case~\eqref{H=0} also \eqref{D2H} is identically satisfied,  while in case \eqref{H=const} it simplifies since $\square H=0$.

\subsection{Ansatz in hypersferical coordinates}

\label{subsec_ans2}

Similar arguments also apply in the coordinate system $(\alpha,F,G)$ of \eqref{pions2}. In particular, if we assume $x^1=G$ to be Killing and $g^{G\hat\mu}=0$ for all coordinates $x^{\hat\mu}$ with $\hat\mu=2,3,4$ (except, possibly, for additional Killing coordinates), then $\square G=0$.
By further setting $\alpha=\pi/2$ (this means that the $SU(2)$ field takes an $S^2$ configuration), \eqref{fequ1} becomes an identity, while \eqref{fequ3} reduces to $\sin\left( 2F\right)\nabla G\cdot\nabla F=0$. If we take $x^2=F$ (not being a spacetime Killing coordinate), then $g^{GF}=0$ (by our previous assumption), so that also \eqref{fequ3} is identically satisfied. In this work we will further assume, for simplicity, that also $x^3$ is a spacetime Killing coordinate.

To summarize, we have defined three spacetime Killing coordinates
\be
 x^1=G , \quad x^2=F , \quad x^3
 \label{ans2_coords}
\ee
and assumed the metric conditions
\be
 g^{GF}=g^{G 4}=0 .
 \label{ans2_metric}
\ee
Since $\alpha=\pi/2$, here one has
\be
	{S}_{\mu \nu }=\left( \nabla _{\mu }F\right)\left( \nabla _{\nu }F\right) +\sin ^{2}F\left( \nabla _{\mu }G\right)\left( \nabla _{\nu }G\right) .
\ee
and the only remaining field equation \eqref{fequ2} reads
\be
 \square F-\frac{1}{2}\sin (2F)\nabla G\cdot\nabla G=0 .
 \label{D2F}
\ee

We will now discuss separately the various ans\"atze described above in sections~\ref{subsec_ans1} and \ref{subsec_ans2}. We emphasize that they provide, in general, only sufficient conditions in order for (at least some of) the matter field equations to be satisfied, and not necessary ones. Indeed, we will construct also a few examples where the above assumptions can be relaxed, giving rise to solutions with less symmetries. On the other hand, it should also be noted that, even when the assumptions on the metric coefficients are satisfied, it is not guaranteed that the remaining field equations \eqref{D2H} or \eqref{D2F} admit a solution. For example, the Kerr metric in Boyer-Lindquist coordinates obeys the assumptions of the above ansatz \eqref{ans2_coords}, \eqref{ans2_metric}, however \eqref{D2F} is not satisfied when there is non-zero angular momentum (see also appendix~\ref{app_kerr}). A similar comment applies also to the C-metric in spherical-type coordinates (cf., e.g., section~14.1.1 of \cite{GriPodbook}) when the acceleration parameter is non-zero. It is therefore non-trivial that some spacetime exist which admit solutions of \eqref{D2H} or \eqref{D2F} (and that, moreover, also allow for the backreaction of the pionic fields to be consistently taken into account).

\section{Toroidal ansatz}

\label{sec_1ai}

The first class of solutions we consider corresponds to the ansatz~\eqref{H=const} of section~\ref{subsec_ans1}. It will lead to toroidal black holes, their extremal (i.e., near-horizon) limits, and their extensions with a NUT parameter or in the Robinson-Trautman and Kundt classes.

In the parametrization \eqref{HAG_def}, let us assume that
\be
 H=\frac{\pi}{4} ,
 \label{H_spec}
\ee
and the remaining pionic fields define two spacetime Killing coordinates
\be
 (x^1,x^2)\equiv(A,G) .
\ee
As explained in section~\ref{subsec_ans1} (case~\eqref{H=const}), assuming
\be
 g^{A 3}=g^{A 4}=0=g^{G 3}=g^{G 4} ,
 \label{ans1_tor}
\ee
ensures that the pionic field equations reduce to \eqref{D2H}, which now gives
\be
 g^{AA}=g^{GG} .
 \label{eqH_1ai}
\ee

With the above assumptions, one can write the metric as
\begin{equation}
 ds^{2}=g^{(1)}_{BC}(x^3,x^4)dx^Adx^B+g^{(2)}_{bc}(x^3,x^4)dx^adx^b ,
\label{metric_generic}
\end{equation}
where $B,C=1,2$ and $b,c=3,4$, while \eqref{eqH_1ai} becomes equivalent to
\be
 g_{AA}=g_{GG} .
 \label{gAA_gGG_dd}
\ee
This case includes a subset of stationary axisymmetric spacetimes, but the two Killing vectors can also be both spacelike. With no loss of generality, a $(x^3,x^4)$ coordinate transformation can be used to set $g^{(2)}_{33}= \pm g^{(2)}_{44}$ (the sign depending on the choice of a time coordinate) and $g^{(2)}_{34}=0$, if desired.

While the pionic field equations are identically satisfied in any spacetime \eqref{metric_generic} with \eqref{eqH_1ai}, it is not a priori obvious that Einstein's equations \eqref{einstein2} can be consistently solved. We will now show that this can be done explicitly at least in some special cases of particular physical interest. Moreover, this will subsequently inspire educated guesswork that will lead to more general solutions of the coupled Einstein-nonlinear $\sigma$-model equations which go beyond the original ansatz.

\subsection{Special case with adapted 2-geometry: toroidal black holes and their extremal limit}

\label{subsec_1a_diag}

For simplicity, let us assume from now on that
\be
 g_{AG}=0 .
\ee
This means that the metric of the 2-space $(A,G)$ is proportional to the tensor \eqref{S_tor} (with \eqref{H_spec}; recall \eqref{gAA_gGG_dd}) and that, to preserve the Lorentzian signature, $g_{AA}>0$.
Two cases need to be analyzed separately.

\subsubsection{Case $dg_{AA}\neq0$: toroidal black hole}

\label{subsubsec_1ai_BH}

In this case we can define coordinates $(x^3,x^4)=(t,r)$ such that
\be
 r^2=g_{AA} .
\ee
For $g_{tt}\neq0$, a new coordinate $t'(t,r)$ can always be defined (without changing $r$) such that $g_{t'r}=0$, i.e., (dropping the primes) the metric takes the form $ds^2=-f(t,r)dt^2+h(t,r)dr^2+r^2(dA^2+dG^2)$. Next, it is not difficult to see (similarly as in the proof of Birkhoff's theorem, cf., e.g., \cite{PleKrabook}) that the solution to Einstein's equations is given (up to normalizing $t$) by
\be
 ds^2=-f(r)dt^2+\frac{dr^2}{f(r)}+r^2(dA^2+dG^2) , \qquad f(r)=-\quattG -\frac{2m}{r}-\frac{\Lambda}{3}r^2 .
 \label{BH_flat}
\ee
This spacetime represents a toroidal black hole in an asymptotically locally AdS spacetime. It was found and studied in \cite{Astorinoetal18}. The (bi-)axionic counterpart of this solutions (dualized to 3-forms) was obtained earlier in \cite{BarCalCha12} (see also \cite{Caldarellietal17}, and \cite{Tanetal18} in higher dimensions). The same metric, but supported by a complex scalar field, was found in \cite{Chenetal10}.

The case $g_{tt}=0$ turns out to be incompatible with Einstein's equations and therefore there are no other solutions in this branch.

\subsubsection{Case $dg_{AA}=0$: AdS$_2\times S^1\times S^1$}

\label{subsubsec_1ai_prod}

Here $g_{AA}=b_0^2$ is a positive constant. Einstein's equations then imply that the 2-space spanned by $(x^3,x^4)$ has constant Gaussian curvature given by $\Lambda$ and that $2b_0^2\Lambda+\kappa K=0$. The metric can thus be reduced to a direct product AdS$_2\times S^1\times S^1$
\be
 ds^2=-|\Lambda|r^2dt^2+\frac{dr^2}{|\Lambda|r^2}+b_0^2(dA^2+dG^2) , \qquad \Lambda=-\frac{\kappa K}{2b_0^2}<0  .
 \label{AdS2_R2}
\ee
This spacetime can also be obtained as the extremal limit of \eqref{BH_flat} (when $\Lambda<0$). It is worth recalling that the same metric also describes an electrovac solution in the standard Einstein-Maxwell theory \cite{PlebHac79} (but with a different value of $\Lambda$; see also \cite{KunLuc13} and references therein for its role as a near-horizon geometry).

\subsection{Extended ansatz I: NUT metric with flat base space}

The assumption that both pionic fields $(A,G)$ are identified with spacetime Killing coordinates can be in some cases relaxed. An interesting example when this occurs is given by the NUT-like metric
\be
 ds^2=-f(r)(dt-2\ell ydx)^2+\frac{dr^2}{f(r)}+(r^2+\ell^2)(dx^2+dy^2) .
 \label{NUT_0}
\ee
It is easy to see that defining
\be
 A=x , \qquad G=y ,
 \label{pions_NUT_0}
\ee
the pionic field equations \eqref{D2H}--\eqref{D2ph} are satisfied identically (even if $\pa_y$ is not a Killing vector). This is true for any choice of $f(r)$, which enables us to take into account also the effect of the backreaction.  Namely, a solution of the Einstein equations \eqref{einstein2} sourced by \eqref{pions_NUT_0} is given by metric~\eqref{NUT_0} with
\be
 f(r)=\frac{-\frac{\Lambda}{3}r^4-2\Lambda \ell^2r^2-2mr+\Lambda\ell^4-\quattG (r^2-\ell^2)}{r^2+\ell^2} .
 \label{NUT_flat}
\ee
For $K=0$ this reduces to the $\Lambda$-NUT solution with a flat base space \cite{NewTamUnt63,Carter68cmp,CahDef68}, while for $\ell=0$ to the toroidal black hole \eqref{BH_flat}. This nutty spacetime can thus also be reinterpreted as an extension of the bi-axionic solution of \cite{BarCalCha12}.

\subsection{Extended ansatz II: Robinson-Trautman solutions}

\label{subsec_RT_flat}

A different extension of the ansatz of section~\ref{subsubsec_1ai_BH} enables one to construct somewhat more general solutions. Namely, since the toroidal black holes \eqref{BH_flat} clearly belong to the well-known Robinson-Trautman class of spacetimes \cite{Stephanibook}, it is natural to study whether one can obtain other pionic solutions in such a family of spacetimes.

Let us consider the general Robinson-Trautman line-element \cite{Stephanibook}
\be
 ds^2=2r^2P^{-2}d\zeta d\bar\zeta-2dudr-2{\cal H}du^2 ,
 \label{RT}
\ee
where $P=P(u,\zeta,\bar\zeta)$, ${\cal H}={\cal H}(u,r,\zeta,\bar\zeta)$. If we identify
\be
 \zeta=\frac{A+iG}{\sqrt{2}} ,
 \label{zeta}
\ee
then the field equations~\eqref{D2H}--\eqref{D2ph} are identically satisfied for any choice of $P$ and ${\cal H}$, since $H$ is constant (even if now $A$, $G$ are generically {\em not} spacetime Killing coordinates).

Einstein’s equations enable one to express the metric function ${\cal H}$ as
\be
 2{\cal H}=\Delta\ln P-\quattG P^2-2r(\ln P)_{,u}-\frac{2m(u)}{r}-\frac{\Lambda}{3}r^2 ,
\ee
up to solving the (modified) Robinson-Trautman equation
\be
 \Delta\Delta\ln P-\quattG \Delta(P^2)+12m(\ln P)_{,u}-4m_{,u}=0 ,
 \label{RT_eq}
\ee
where $\Delta\equiv 2P^2\pa_\zeta\pa_{\bar\zeta}$ is the Laplace operator in the 2-space $(\zeta,\bar\zeta)$.

The Weyl tensor component
\be
 \Psi_2=-\frac{m}{r^3}-{\frac{\kappa KP^2}{12r^2}} ,
\ee
shows that these spacetimes are of Petrov type II or D (since $\Psi_0=\Psi_1=0$). It also shows that the pionic term produces a ``non-standard'' (slower) fall-off near infinity.

This family of solutions includes the static black hole~\eqref{BH_flat} for the special choice $P=1$, $m=$const (up to a simple coordinate transformation), but it also describes time-dependent solutions. When \eqref{RT_eq} is {\em not} satisfied, the energy-momentum contains an additional pure radiation contribution $T_{uu}$ (with $\Delta\Delta\ln P-\quattG \Delta(P^2)+12m(\ln P)_{,u}-4m_{,u}>0$ by the null energy condition). The special case $P=1$ gives radiating Vaidya-type solutions, as noticed in \cite{BarCalCha12} in the axionic case.

It would be interesting to find other explicit solutions of~\eqref{RT_eq} and study their physical properties. However, finding a general solution to \eqref{RT_eq} is a highly non-trivial problem also in the vacuum case $K=0$ (cf. \cite{Stephanibook} and references therein), therefore we leave this for possible future investigations.

\subsection{Extended ansatz III: Kundt solutions}

\label{subsec_K_flat}

We now consider an extension of the ansatz of section~\ref{subsubsec_1ai_prod}. Since the latter spacetime belongs to the Kundt class, let us start here with the general Kundt line-element \cite{Stephanibook}
\be
 ds^2=2P^{-2}d\zeta d\bar\zeta-2du(dv+Wd\zeta+\bar Wd\bar\zeta+{\cal H}du) ,
 \label{K}
\ee
where again we have defined $(\zeta,\bar\zeta)$ using~\eqref{zeta}, and $P_{,v}=0$, $W=W(u,v,\zeta,\bar\zeta)$, ${\cal H}={\cal H}(u,v,\zeta,\bar\zeta)$. (Since $\zeta$ is dimensionless, here $P$ has the dimension of an inverse length.)

The pionic field equations \eqref{D2ga}, \eqref{D2ph} are satisfied iff
\be
 W_{,v}=0 ,
 \label{Wv}
\ee
which means that the Kundt vector field $\pa_v$ is recurrent, while \eqref{D2H} is satisfied identically (recall \eqref{H_spec}). Using the Ricci tensor components given in \cite{Stephanibook} with \eqref{Wv}, it is easy to see that the Einstein equations give
\be
 {\cal H}=-\frac{1}{2}\Lambda v^2+{\cal H}^{(1)}(u,\zeta,\bar\zeta)v+{\cal H}^{(0)}(u,v,\zeta,\bar\zeta) ,
 \label{H_K}
\ee
along with the following differential equations\footnote{An additional equation coming from the term linear in $v$ in the Einstein equation of boost weight $-2$ can be shown to be a consequence of \eqref{DeltaP_kundt}, \eqref{H1_kundt} -- therefore we do not display it (cf. \cite{Stephanibook} for a similar comment in the electrovacuum case).} for the unknown functions $P$, ${\cal H}^{(1)}$, ${\cal H}^{(0)}$ and $W$
\beqn
 & & \Delta\ln P=\Lambda+\quattG P^2 , \label{DeltaP_kundt} \\
 & & {\cal H}^{(1)}_{,\zeta}+(P^2W)_{,\zeta\bar\zeta}-2(\ln P)_{,\bar\zeta}(P^2W)_{,\zeta}-\mu_{,\zeta}=\quattG P^2W , \label{H1_kundt}\\
 & & \Delta\left({\cal H}^{(0)}+P^2W\bar W\right)-2(P^2\bar W)_{,\bar\zeta}(P^2W)_{,\zeta}-2\left[\mu_{,\zeta} P^2\bar W+\mu_{,\bar\zeta}P^2W+\mu_{,u}+\mu {\cal H}^{(1)}+\mu^2\right] \nonumber \\
 & & \qquad\qquad\qquad\qquad\qquad\qquad\qquad\qquad\qquad\qquad =\ottG KP^4W\bar W ,
\eeqn
where
\be
 \mu\equiv\frac{1}{2}P^2(\bar W_{,\zeta}+W_{,\bar\zeta})-(\ln P)_{,u} ,
 \label{mu}
\ee
is one of the Newman-Penrose coefficients.

These spacetimes are of Petrov type II or D, since $\Psi_0=\Psi_1=0$, whereas
\be
 \Psi_2=-\frac{1}{6}(\Delta\ln P+\Lambda) ,
 \label{psi2_K}
\ee
which is necessarily non-zero in view of \eqref{DeltaP_kundt} (except in the vacuum limit with $\Lambda=0=\Delta\ln P$).
Note also that $\Delta\ln P$ is the Gaussian curvature of the 2-surfaces of constant $u$ and $v$ (``wave surfaces'') \cite{Stephanibook}, which is not a constant in general because of the pionic term in \eqref{DeltaP_kundt} (as opposed to the ($\Lambda$-)vacuum case \cite{Stephanibook,GriPodbook}).

The only possible case with wave surfaces of constant curvature occurs when $P$ is assumed to be a non-zero constant $b_0^{-1}$ such that $\Lambda+\quattG b_0^{-2}=0$ (the wave surfaces are thus flat). With the additional simplifying assumption $W=0$, one obtains the special solution (after gauging away the function ${\cal H}^{(1)}(u)$, cf. \cite{Stephanibook})
\be
 ds^2=2b_0^2d\zeta d\bar\zeta-2dudv+\left[\Lambda v^2+f(u,\zeta)+\bar f(u,\bar\zeta)\right]du^2  \qquad (\Lambda=-\quattG b_0^{-2}) .
\ee
This solution describes a non-expanding gravitational wave propagating in the AdS$_2\times S^1\times S^1$ pionic spacetime \eqref{AdS2_R2}, to which it reduces for $f=0$ (in slightly different coordinates -- recall also \eqref{zeta}). Similar solutions in the Einstein-Maxwell theory were found in \cite{GarAlvar84,Khlebnikov86} and further studied in \cite{OrtPod02,PodOrt03}. The more general case $W\neq0$ gives rise to gyratonic solutions similar to those analyzed (in the Einstein-Maxwell theory) in \cite{Kadlecovaetal09} and may deserve a separate investigation.

\section{Spherical ansatz}

\label{sec_3b}

We describe now the second class of solutions, which corresponds to the ansatz of section~\ref{subsec_ans2}. It will lead to spherical black holes, their extremal limits, and their extensions with a NUT parameter or in the Robinson-Trautman and Kundt classes.

In the parametrization~\eqref{pions2}, we assume
\be
 \alpha=\frac{\pi}{2} ,
\ee
and define the spacetime coordinates
\be
 (x^1,x^2,x^3,x^4)\equiv(G,F,t,r) ,
\ee
where $(G,t)$ are assumed to be Killing coordinates. We further require
\be
 g^{GF}=0=g^{Gr} .
 \label{gGF_gGr}
\ee
As explained in section~\ref{subsec_ans2}, in any such spacetime the pionic field equations reduce to the single equation \eqref{D2F}. Some coordinate freedom can be employed to simply the line-element depending on whether $g^{FF}=0$ or not.

If $g^{FF}\neq 0$, a redefinition of $t\mapsto t'=t+T(r,F)$, $r\mapsto r'=R(r,F)$ can be used to set
\be
 g^{tF}=0=g^{rF} .
 \label{gGF_gGr_transf1}
\ee
In terms of the covariant metric, \eqref{gGF_gGr} and \eqref{gGF_gGr_transf1} read
\be
 g_{tF}=g_{rF}=g_{FG}=0 , \qquad g_{tt}g_{rG}-g_{tG}g_{tr}=0 ,
 \label{3b_covariant}
\ee
i.e.,
\be
 ds^2=b^2(r,F)dF^2+g_{ij}(r,F)dx^i dx^j ,
 \label{metric3b}
\ee
where $i,j=1,3,4$, and the 3-metric $g_{ij}$ is constrained by the second of \eqref{3b_covariant} (for definiteness we have assumed that $\pa_F$ is spacelike). Recall also that one still needs to solve the pionic equation \eqref{D2F}.

If $g^{FF}=0$, we can redefine $t\mapsto t'=t+T(r,F)$ such that also\footnote{Except when $g^{rF}=0$. However, this case can be neglected since one has $\Box F=0$, for which \eqref{D2F} admits no solution ($\det g\neq0$ requiring $g^{GG}\neq0$).}
\be
 g^{tF}=0 ,
\ee
so that the covariant metric is constrained by
\be
 g_{tr}=g_{rr}=g_{rG}=0 , \qquad g_{tt}g_{FG}-g_{tG}g_{tF}=0 ,
 \label{3b_covariant_special}
\ee
and necessarily $g_{rF}\neq0$. In this case, requiring $\det g<0$ implies $g_{tt}g_{GG}>0$ and from \eqref{D2F} one obtains
\be
 g_{rF}=\frac{(g_{tt}g_{GG}-g_{tG}^2)_{,r}}{g_{tt}\sin(2F)} .
\ee

\subsection{Special diagonal subcase with adapted 2-geometry ($g^{FF}\neq 0$): spherical black holes and their extremal limit}

\label{subsec_sphericalBH}

Let us consider the ansatz \eqref{metric3b} and further assume that the 3-metric $g_{ij}$ is diagonal (which guarantees that \eqref{3b_covariant} is satisfied) and, in addition,
\be
 b_{,F}=0 , \qquad g_{GG}=b\sin^2F .
\ee

From \eqref{D2F} one immediately obtains $g_{rr}=c_1(r)/g_{tt}$, while a linear combination of the Einstein equations $(FF)$ and $(GG)$ gives $g_{tt,F}=0$.

\subsubsection{Case $b_{,r}\neq0$: spherical black hole}

\label{subsubsec_S2_BH}

In this case we can redefine $r\mapsto r'(r)$ such that $r'=b(r)$. Integrating Einstein's equations then yields (dropping the prime and after a constant rescaling of $t$)
\be
 ds^2=-f(r)dt^2+\frac{dr^2}{f(r)}+r^2(dF^2+\sin^2FdG^2) , \qquad f(r)=(1-\kappa K)-\frac{2m}{r}-\frac{\Lambda}{3}r^2 .
 \label{BH_S2}
\ee
For $\Lambda=0$, this solution was obtained in \cite{BarVil89,Gibbons91} and describes a global monopole inside a black hole (in the $\sigma$-model approximation) -- it is not asymptotically flat but contains a solid angular deficit. The generalization with $\Lambda\neq0$ was obtained in \cite{GueRab91} and with a Skyrme term in \cite{CanMae13}.

\subsubsection{Case $b_{,r}=0$: Levi-Civita--Bertotti--Robinson and Nariai solutions (A)dS$_2\times S^2$ and $\mathbb{R}^1_1\times S^2$}

\label{subsubsec_S2_extremal}

Here $g_{FF}=b_0^2=$const, while a redefinition of $r$ allows us to set $g_{rr}=1/g_{tt}$. Einstein's equations then lead to
\be
 ds^2=-f(r)dt^2+\frac{dr^2}{f(r)}+b_0^2(dF^2+\sin^2FdG^2) , \qquad f(r)=-\Lambda r^2+c_1r+c_2 , \qquad \Lambda=\frac{1-\kappa K}{b_0^2}  .
 \label{BR}
\ee
This is a direct product of two 2-spaces of constant Gaussian curvature equal to $\Lambda$ and $1/b_0^2$, respectively. Depending on the sign of $1-\kappa K$, this is thus an (A)dS$_2\times S^2$ or $\mathbb{R}^1_1\times S^2$ spacetime, similar to the well-known electrovac solutions of \cite{LeviCivita17BR,Nariai51,Bertotti59,Robinson59,PlebHac79}. This metric can also be obtained as the extremal limit of \eqref{BH_S2} (when $1-\kappa K\neq0$). It is conformally flat iff the constants are fine tuned as $\quattG =1$.

\subsection{NUT metric with spherical base space}

Although the static black hole of section~\ref{subsubsec_S2_BH} cannot be extended to a pionic version of the Kerr black hole (as discussed in appendix~\ref{app_kerr}), it does admit a generalization to a stationary NUT spacetime with a spherical base space. To see this, let us consider the line-element
\be
 ds^2=-f(r)\left(dt+4\ell\sin^2\frac{\theta}{2}d\phi\right)^2+\frac{dr^2}{f(r)}+(r^2+\ell^2)(d\theta^2+\sin^2\theta d\phi^2) .
 \label{NUT_1}
\ee
This clearly has the form of ansatz~\eqref{metric3b}, provided one takes a pionic field configuration with
\be
 F=\theta , \quad G=\phi .
 \label{pions_NUT_1}
\ee
Furthermore, using \eqref{NUT_1}, \eqref{pions_NUT_1} it is easy to verify that also the remaining pionic field equation~\eqref{D2F} is satisfied identically, for any choice of $f(r)$. This freedom enables us to take into account also the effect of the backreaction on the metric. Namely, a solution of the Einstein equations \eqref{einstein2} sourced by \eqref{pions_NUT_1} (recall that $\alpha=\pi/2$) is given by metric~\eqref{NUT_1} with
\be
 f(r)=\frac{-\frac{\Lambda}{3}r^4+r^2(1-2\Lambda \ell^2)-2mr-\ell^2(1-\Lambda\ell^2)-\ottG K(r^2-\ell^2)}{r^2+\ell^2} .
\ee
This solution was already obtained in \cite{CanMae13} (including a Skyrme term). For $\ell=0$ it reduces to the black hole \eqref{BH_S2}, while for $K=0$ it gives the vacuum Taub-NUT solution \cite{Taub51,NewTamUnt63} with $\Lambda$ \cite{Carter68cmp,CahDef68}

\subsection{Extended ansatz I: Robinson-Trautman solutions}

\label{subsec_RT_spher}

Similarly as in section~\ref{subsec_RT_flat}, it is natural to consider an extension of the spherical black holes \eqref{BH_S2} to the more general Robinson-Trautman class. The line-element is still given by \eqref{RT}, but now with the identification
\be
 \zeta=\sqrt{2}e^{iG}\tan\frac{F}{2} .
 \label{zeta_2}
\ee
Taking $\alpha=\pi/2$ (as before), the field equations~\eqref{fequ1}--\eqref{fequ3} are identically satisfied for any choice of $P$ and ${\cal H}$ (even if now the spacetime does not possess any Killing vectors, generically).

Defining
\be
  Q\equiv 1+\tan^2\frac{F}{2}=1+\frac{1}{2}\zeta\bar\zeta ,
	\label{Q}
\ee
Einstein’s equations give
\be
 2{\cal H}=\Delta\ln P-\ottG KP^2Q^{-2}-2r(\ln P)_{,u}-\frac{2m(u)}{r}-\frac{\Lambda}{3}r^2 ,
\ee
along with the modified Robinson-Trautman equation (as before $\Delta\equiv 2P^2\pa_\zeta\pa_{\bar\zeta}$)
\be
 \Delta\Delta\ln P-\ottG K\Delta(P^2Q^{-2})+12m(\ln P)_{,u}-4m_{,u}=0 .
 \label{RT_eq_2}
\ee

As in section~\ref{subsec_RT_flat}, it is easy to argue that these spacetimes can only be of Petrov type II or D, with a $1/r^2$ fall-off of $\Psi_2$ due to the pionic term.

The static black holes \eqref{BH_S2} are recovered in the special case $P=Q$, $m=$const. When \eqref{RT_eq_2} is {\em not} satisfied, the energy-momentum contains also a pure radiation term $T_{uu}$ -- in particular, for $P=Q$, $m_{,u}\neq0$ one a pionic extension of the Vaidya solution \cite{Stephanibook}. As observed in section~\ref{subsec_RT_flat}, it would be interesting to study this class of solutions more systematically, which would deserve a separate investigation.

\subsection{Extended ansatz II: Kundt solutions}

\label{subsec_K_spher}

Following the ideas of section~\ref{subsec_K_flat}, we can similarly extend the ansatz of section~\ref{subsubsec_S2_extremal} to more general Kundt solutions. One starts again with the line-element \eqref{K}, but now with the identification \eqref{zeta_2}. With $\alpha=\pi/2$, the field equations~\eqref{fequ1}--\eqref{fequ3} are identically satisfied provided \eqref{Wv} holds (while $P$ and ${\cal H}$ remain arbitrary). Einstein's equations again imply \eqref{H_K}, but now with
\beqn
 & & \Delta\ln P=\Lambda+\ottG KP^2Q^{-2} , \label{DeltaP_kundt_2} \\
 & & {\cal H}^{(1)}_{,\zeta}+(P^2W)_{,\zeta\bar\zeta}-2(\ln P)_{,\bar\zeta}(P^2W)_{,\zeta}-\mu_{,\zeta}=\ottG KP^2Q^{-2}W , \label{H1_kundt_2}\\
 & & \Delta\left({\cal H}^{(0)}+P^2W\bar W\right)-2(P^2\bar W)_{,\bar\zeta}(P^2W)_{,\zeta}-2\left[\mu_{,\zeta} P^2\bar W+\mu_{,\bar\zeta}P^2W+\mu_{,u}+\mu {\cal H}^{(1)}+\mu^2\right] \nonumber \label{H0_kundt_2} \\
 & & \qquad\qquad\qquad\qquad\qquad\qquad\qquad\qquad\qquad\qquad =\sedG KP^4Q^{-2}W\bar W ,
\eeqn
where \eqref{mu} and \eqref{Q} are understood.

The Weyl scalar $\Psi_2$ is again given by \eqref{psi2_K}. These spacetimes are thus generically of Petrov type II, however, the types III and N are now also possible (as opposed to the case of section~\ref{subsec_K_flat}). First, imposing $\Psi_2=0$ implies that the 2-space $(\zeta,\bar\zeta)$ have constant Gaussian curvature $-\Lambda$. Then, compatibility with \eqref{DeltaP_kundt_2} requires $\Lambda<0$, $P=\sqrt{-\Lambda}Q$ and the fine-tuning $\quattG =1$ (and one has to solve \eqref{H1_kundt_2}, \eqref{H0_kundt_2} in a simplified form). A simple example (of type N) occurs for $W=0={\cal H}^{(1)}$, i.e.,
\be
 ds^2=2|\Lambda|^{-1}Q^{-2}d\zeta d\bar\zeta-2dudv+\left[\Lambda v^2+f(u,\zeta)+\bar f(u,\bar\zeta)\right]du^2  \qquad (\Lambda<0 , \quad \quattG =1) .
\ee

This solution describes a non-expanding gravitational wave propagating in a conformally flat AdS$_2\times S^2$ pionic background, to which it reduces for $f=0$  (the latter is contained in the ``near-horizon'' solution \eqref{BR}, up to a simple coordinate transformation).
See \cite{Khlebnikov86,Ortaggio02,OrtPod02,PodOrt03} for similar solutions in the Einstein-Maxwell theory.

\section{Circular ansatz}

\label{sec_circular}

Let us turn to the ansatz~\eqref{H=0} of section~\ref{subsec_ans1}. It will lead to black string-type solutions.

In the parametrization \eqref{HAG_def}, let us assume that
\be
 H=0 ,
\ee
and the remaining pionic fields define two spacetime Killing coordinates
\be
 (x^1,x^2)\equiv(A,G) .
\ee
As discussed in section~\ref{subsec_ans1} (case~\eqref{H=0}), assuming $g^{A 3}=g^{A 4}=0=g^{G 3}=g^{G 4}$, ensures that the pionic field equations are identically satisfied. One thus only needs to solve Einstein's equations. We will discuss this here only in one special case, which appears however of some interest. Moreover, thanks to the comments in section~\ref{subsec_ans1}, various solutions available in the literature on Einstein gravity coupled to a massless scalar field can be reinterpreted as nonlinear sigma-model solutions within this ansatz.

\subsection{Special static case: black strings}

Let us further assume here that the spacetime is static and define coordinates $(x^1,x^2,x^3,x^4)\equiv(A,G,t,r)$. Because of the form of the energy-momentum tensor (cf. section~\ref{subsec_ans1}), it is natural to consider two possible different ans\"atze where the coordinates $A$ and $G$ play a different role, namely $ds^2=-f(r)dt^2+dr^2/f(r)+r^2dA^2+b_0^2dG^2$ and $ds^2=-f(r)dt^2+dr^2/f(r)+r^2dG^2+b_0^2dA^2$, where $b_0$ is a constant (the case when $g_{AA}$ and $g_{GG}$ are both constant turns out to admit no solution here).

In the first case, integrating Einstein's equations implies $\Lambda=0$ and leads to the solution
\be
 ds^2=\kappa K\ln\frac{r}{r_0}dt^2+\frac{dr^2}{-\kappa K\ln\frac{r}{r_0}}+r^2dA^2+b_0^2dG^2  \qquad (\Lambda=0) ,
 \label{string1}
\ee
where $r_0$ is an integration constant. This is the direct product of a (2+1)-dimensional electrostatic solution found in \cite{DesMaz85,Melvin86,GotSimAlp86} with a circle or arbitrary radius. It represents an object with a $S^1\times S^1$ horizon located at $r=r_0$ and a static region for $0<r<r_0$, which ends at a timelike curvature singularity at $r=0$ (since ${\cal R}=\kappa K/r^2$), thus resembling the inner region of the Reissner-Nordstr\"om spacetime. If reinterpreted as a scalar field solution (see section~\ref{subsec_ans1}), the range of the ``angular'' coordinates becomes arbitrary and the same solution can thus also describe a black string or a black 2-brane. It can be lifted to higher dimensions by simply taking the direct product with a Ricci-flat space (see \cite{Giacominietal18} for a recent discussion of a similar construction). It should be observed that a black string solution similar to~\eqref{string1} was obtained and its thermodynamics discussed in \cite{Astorinoetal18_prd} -- that solution however has $\Lambda<0$ and in fact corresponds to a different ansatz for the $SU(2)$ matter field.

The second solution is given by
\be
 ds^2=-\left(-\mu-\frac{\Lambda}{2}r^2\right)dt^2+\frac{dr^2}{-\mu-\frac{\Lambda}{2}r^2}+r^2dG^2+b_0^2dA^2  \qquad  \Lambda=-\frac{\kappa K}{b_0^2}<0 ,
\ee
where again $\mu$ is an integration constant which we now assume to be positive (in order to have an event horizon). Note that here the cosmological constant is necessarily negative. This is the direct product of the vacuum BTZ black hole \cite{BTZ} with a circle, whose radius fixes the value of $\Lambda$. The same solution was found in \cite{CisOli18,CisCorrdel19} in the context of general relativity coupled to a scalar field (cf. the comments in section~\ref{subsec_ans1}), in which case the range of $A$ is arbitrary (while $G$ has still to be periodic in order for sections at constant $A$ to describe a BTZ black hole). It can be lifted to higher dimensions provided an additional scalar field is introduced for each extra dimension (so as to support the cosmological constant) \cite{CisOli18}. Different BTZ black strings were discussed in \cite{EmpHorMye00II}: those are vacuum solution and the fourth dimension is necessarily warped.

\section{Hyperspherical ansatz}

\label{sec_hypers}

Let us discuss now the last class of solutions considered in this paper, which corresponds to the ansatz~\eqref{H_generic} of section~\ref{subsec_ans1}. It will lead to cosmological solutions (but also to their stationary ``counterparts'') which, as mentioned in section~\ref{sec_system}, have a non-zero topological charge (see \cite{AyoCanZan16,Canforaetal19} and references therein for related comments).

In the parametrization \eqref{HAG_def}, let us take
\be
 (x^1,x^2,x^3,x^4)\equiv(A,G,t,H) ,
\ee
where $(A,G)$ are Killing coordinates. As discussed in section~\ref{subsec_ans1} (case~\eqref{H_generic}), assuming
\be
 g^{AH}=g^{At}=0=g^{GH}=g^{Gt} ,
\ee
the field equations reduce to \eqref{D2H}. The metric has the form~\eqref{metric_generic} (but without \eqref{gAA_gGG_dd}).

\subsection{Special case with adapted 3-geometry: LFRW cosmologies and Einstein's static universe}

\label{subsec_1aiii_LFRW}

Let us assume
\be
 g_{AA}=f(t,H)\cos^2H , \qquad g_{GG}=f(t,H)\sin^2H , \qquad g_{AG}=0 ,  \qquad  f(t,H)\equiv g_{HH} ,
 \label{1aiii_adapted}
\ee
where $f$ is a positive function. This means that the metric of the 3-spaces at constant $t$ is proportional to the tensor \eqref{S_H}.
 Similarly as in section~\ref{subsubsec_1ai_BH}, if $g_{tt}\neq0$ we can redefine $t$ such that $g_{tH}=0$.\footnote{The case $g_{tt}=0$ turns out to be forbidden, as can be seen by first setting $g_{HH}=0$ by a redefinition of $t$, and then considering the Einstein equation $(tt)$ with \eqref{D2H}.} Eq.~\eqref{D2H} then gives (up to normalizing $t$)
\be
 g_{tt}=-\frac{1}{f(t,H)} ,
\ee
and all the matter field equations are now solved. Integrating the Einstein equation $(tH)$ gives $f=[f_1(t)+f_2(H)]^2$, but a linear combination of the equations $(tt)$, $(HH)$ and $(AA)$ then reveals that $f_2$ must be a constant and therefore $f=f(t)$. Solving the remaining Einstein equation one arrives at the final metric
\beqn
  & & ds^2=-\frac{dt^2}{f(t)}+f(t)(dH^2+\cos^2 H dA^2+\sin^2 H dG^2) , \nonumber \\
  & & f(t)=\frac{\Lambda}{3}t^2+c_1t+c_2 , \qquad \kappa K=\frac{3c_1^2-4\Lambda c_2}{6}+2 . \label{LFRW}
\eeqn

This metric clearly describes a LFRW cosmology (and is thus conformally flat) with spherical spatial sections. The pionic matter acts  effectively as a (non-tilted) {\em perfect fluid} with 4-velocity $\sqrt{f}\pa_t$ and energy density $\rho=3K/(2f)$ and negative pressure $p=-K/(2f)$, thus satisfying a barotropic equation of state $p=(\gamma-1)\rho$ with $\gamma=2/3$. We can expect that, since $\rho+3p=0$, accelerated expansion occurs only when $\Lambda>0$ (as we indeed demonstrate explicitly below). A constant shift of $t$ (under which the quantity $3c_1^2-4\Lambda c_2$ in \eqref{LFRW} is invariant) can always be used to get rid of one of the integration constants $c_1$ and $c_2$ (cf., e.g., \cite{OrtPraPra11}). One thus arrives at various canonical forms of the metric with either
\be
 c_1^2=0, \quad c_2=-3(\kappa K-2)/(2\Lambda) , \qquad \mbox{or } \qquad c_1^2=2(\kappa K-2), \quad c_2=0 .
\ee
Since one must ensure $f(t)>0$, eq.~\eqref{LFRW} implies that $\kappa K-2>0$ if $\Lambda<0$ and $\kappa K-2\ge0$ if $\Lambda=0$. Introducing the standard cosmic time by $d\tau=f^{-1/2}dt$ and depending on the possible signs of $\Lambda$ and $\kappa K-2$, similarly as in \cite{OrtPraPra11} one finds:

\begin{enumerate}[(i)]

\item for $\Lambda<0$, $f=c_2\cos^2\left(\sqrt{\frac{|\Lambda|}{3}}\tau\right)$, $\kappa K-2>0$

\item for $\Lambda>0$,

	\begin{enumerate}[a)]
		\item $f=c_2\cosh^2\left(\sqrt{\frac{\Lambda}{3}}\tau\right)$ if $\kappa K-2<0$ (approaching de~Sitter spacetime in the vacuum limit $K\to0$)
		\item $f=|c_2|\sinh^2\left(\sqrt{\frac{\Lambda}{3}}\tau\right)$ if $\kappa K-2>0$
		\item $f=l_0^2e^{\pm2\sqrt{\frac{\Lambda}{3}}\tau}$ if $\kappa K-2=0$ ($l_0$ is an arbitrary length scale)
	\end{enumerate}
	
\item for  $\Lambda=0$,

	\begin{enumerate}[a)]
		
	\item $f=\frac{1}{4}c_1^2\tau^2$, if $\kappa K-2>0$
	\item $f=c_2>0$ if $\kappa K-2=0$, so that the metric becomes
\be
 ds^2=-dt^2+c_2(dH^2+\cos^2 H dA^2+\sin^2 H dG^2)  \qquad (\Lambda=0=\kappa K-2) ,
 \label{Einst_static}
\ee
i.e., an Einstein's static universe $\mathbb{R}\times S^3$ with an arbitrary radius $\sqrt{c_2}$, possessing an extra Killing vector $\pa_t$.

\end{enumerate}

\end{enumerate}

Since $\mathcal{S}=3/f(\tau)$, it is easy to see (recall \eqref{einstein2}) that zeros of $f$ in \eqref{LFRW} correspond to curvature singularities (while the metric is otherwise regular), which occur in the above solutions iff $\kappa K-2>0$. The only oscillating cosmology is obtained for $\Lambda<0$. Solutions~\eqref{LFRW}, \eqref{Einst_static} are well-known in the context of perfect fluid LFRW cosmologies \cite{Harrison67}. The non-linear $\sigma$-model Einstein universe~\eqref{Einst_static} was also found in \cite{Canforaetal17}. Similar solutions when a Skyrme term is added to the theory have been discussed recently in \cite{AyoCanZan16,Pavluchenko16,Canforaetal19} and present different features. In particular, the corresponding Einstein universe differs from \eqref{Einst_static} since it requires $\Lambda\neq0\neq\kappa K-2$ and the radius of the $S^3$ is uniquely fixed by the constants of the theory \cite{AyoCanZan16}. On the other hand, for $\Lambda=0$ there are additional, qualitatively different time-dependent solutions \cite{Pavluchenko16}.

\subsection{Extended ansatz: Bianchi IX cosmologies}

It is worth observing that the solutions considered in section~\ref{subsec_1aiii_LFRW} can be extended to a larger class of spatially homogeneous but anisotropic models, i.e., Bianchi~IX cosmologies. Their line-element can be written as \cite{Taub51} (cf. also, e.g., \cite{RyaShebook,Stephanibook})
\be
 ds^2=-d\tau^2+g_{ij}(\tau)\bm{\omega}^i\bm{\omega}^j \qquad (i,j=1,2,3) ,
 \label{BianchiIX}
\ee
where
\beqn
 & & \bm{\omega}^1=-\sin\chi \,d\theta+\sin\theta\cos\chi \, d\psi \\ \nonumber
 & & \bm{\omega}^2=\cos\chi \,d\theta+\sin\theta\sin\chi \, d\psi \\
 & & \bm{\omega}^3=d\chi+\cos\theta\,d\psi \nonumber ,
\eeqn
define an $SO(3)$ invariant basis. Following \cite{Canforaetal19} (with a slightly different notation), we take the pionic fields \eqref{HAG_def} as
\be
	2H=\theta , \quad 2A=\chi+\psi , \quad 2G=-\chi+\psi .
	\label{euler}
\ee
(The special case of section~\ref{subsec_1aiii_LFRW} is recovered for $g_{ij}(\tau)=\frac{1}{4}f(\tau)\delta_{ij}$.) This means that the invariant 1-forms $R_{\mu }^{i}$ (eq.~\eqref{Ri_2}) coincide with the $\bm{\omega}^i$ (up to a constant rescaling; cf. also appendix~A of \cite{Canforaetal19}) and guarantees that the field equations~\eqref{nonlinearsigma1} are automatically satisfied in the spacetime~\eqref{BianchiIX}.\footnote{This was observed in \cite{Canforaetal19} in the presence of an additional Skyrme term in the theory. Let us note that, alternatively, this also follows easily from the fact that the $\bm{\omega}^i$ form an invariant basis and the dual basis vectors $\e_i$ are linear combinations of the three Killings vectors $\x_i$ \cite{RyaShebook}. For example, writing $\e_i=a\x_1+b\x_2+c\x_3$ and requiring $[\x_1,\e_i]=0$, using $[\x_1,\x_2]=\x_3$ (etc.) one obtains $\nabla_{\x_1}a=0$, and so on with $\x_2$ and $\x_3$, so that eventually $\mbox{div}\,\e_i=0$.}

One thus needs to study only Einstein's equations. In the analysis of \cite{Canforaetal19} it was assumed that~\eqref{BianchiIX} is diagonal. Here we complete this by showing that indeed there is no loss of generality in making this assumption. With~\eqref{S_H} and \eqref{BianchiIX}--\eqref{euler}, the energy-momentum tensor \eqref{timunu2} can be written as
\begin{equation}
	T_{\mu \nu }=K\left[\frac{\mathcal{S}}{2}u_\mu u_\nu+\left(\mathcal{S}_{\mu \nu}-\frac{\mathcal{S}}{3}h_{\mu \nu}\right)-\frac{\mathcal{S}}{6}h_{\mu \nu }\right] ,
 \label{timunu_IX}
\end{equation}
where
\be
 \bm{u}=d\tau , \qquad \bm{h}=g_{ij}(\tau)\bm{\omega}^i\bm{\omega}^j .
\ee
The pionic matter thus acts as an {\em anisotropic fluid} with $\rho=KS/2$, $p=-KS/6$, anisotropic stress $\pi_{\mu \nu }=K\left(\mathcal{S}_{\mu \nu}-\frac{\mathcal{S}}{3}h_{\mu \nu}\right)$ and with vanishing heat flux. Moreover, $\bm{u}$ is clearly a Ricci eigenvector (cf. eq.~\eqref{einstein2}). Thanks to the results of \cite{EllMac69,MacSteSch70} (cf. also the comments in \cite{MacCallum73}), without loss of generality one can thus take $g_{ij}(\tau)$ in \eqref{BianchiIX} to be diagonal, i.e.,
\be
 ds^2=-d\tau^2+a^2(\tau)\bm{\omega}^1\bm{\omega}^1+b^2(\tau)\bm{\omega}^2\bm{\omega}^2 +c^2(\tau)\bm{\omega}^3\bm{\omega}^3  .
 \label{BianchiIX_diag}
\ee
This reduces Einstein's equations to three dynamical equations for the scale factors $a$, $b$, $c$ and one constraint which can be found in \cite{Canforaetal19} (setting the Skyrme coupling constant $\lambda=0$ therein). The analysis of such a system for general Bianchi IX models is very complicated (even in vacuum, cf., e.g., \cite{Taub51,EllMac69,MacCallum73,RyaShebook,Stephanibook}) and goes beyond the scope of this paper. Let us only observe that ref.~\cite{Canforaetal19} derived the corresponding mini-superspace action and discussed the integrability properties of the system.

One might hope to find explicit analytic solutions by making some simplifying assumptions, such as a factorized, overall $\tau$-dependence in \eqref{BianchiIX_diag}. However, easy integration reveals that, in this case, the only solutions are the LFRW ones discussed in section~\ref{subsec_1aiii_LFRW}. In particular, and in contrast to the Skyrme case \cite{Canforaetal19}, the only static solution is thus given by the conformally flat Einstein universe \eqref{Einst_static}. However, different, non-conformally flat solutions are available up to suitable analytic continuations, as we discuss in section \ref{subsec_stationary_Bianchi} below.

\subsection{Special stationary Bianchi IX solutions}

\label{subsec_stationary_Bianchi}

Let us modify the Bianchi~IX ansatz~\eqref{BianchiIX_diag} by taking $\tau=iy$, $a^2=b^2=\frac{1}{4}f$, $c^2=-\frac{1}{4}fq^2$ ($q^2$ is a positive constant), i.e.,
\beqn
  & & ds^2=dy^2+\frac{1}{4}f(y)\left[d\theta^2+\sin^2\theta d\psi^2-q^2(d\chi+\cos\theta d\psi)^2\right] , \nonumber \\
\eeqn
still with the pionic fields~\eqref{euler}.

From Einstein's equations one immediately obtains
\be
 q^2=\frac{1}{4}\kappa K ,
\ee
Then, proceeding as in section~\ref{subsec_1aiii_LFRW} one arrives at the following solutions:
\begin{enumerate}[(i)]

\item for $\Lambda<0$,

	\begin{enumerate}[a)]
		\item $f=c_2\cosh^2\left(\sqrt{\frac{|\Lambda|}{3}}y\right)$ if $\kappa K-8>0$
		\item $f=|c_2|\sinh^2\left(\sqrt{\frac{|\Lambda|}{3}}y\right)$ if $\kappa K-8<0$
		\item $f=l_0^2e^{\pm2\sqrt{\frac{|\Lambda|}{3}}y}$ if $\kappa K-8=0$ ($l_0$ is an arbitrary length scale)
	\end{enumerate}

\item for $\Lambda>0$, $f=c_2\cos^2\left(\sqrt{\frac{\Lambda}{3}}y\right)$, $\kappa K-8<0$

\item for  $\Lambda=0$,

	\begin{enumerate}[a)]
		
	\item $f=\frac{1}{4}c_1^2y^2$, if $\kappa K-8<0$
	\item $f=l_0^2$ if $\kappa K-8=0$, in which case $\pa_y$ is an additional Killing vector.

\end{enumerate}

\end{enumerate}

In the above equations the costants $c_2$ and $c_1$ are given by
\be
 c_2=-\frac{3}{4}\frac{\kappa K-8}{\Lambda} , \qquad c_1^2=-(\kappa K-8) .
\ee

In all cases the Killing vector $\pa_\chi$ is timelike, therefore these spacetimes are stationary. It is also interesting to observe that they are {\em not} conformally flat, as opposed to those of section~\ref{subsec_1aiii_LFRW}. Since $C_{\mu\nu\rho\sigma}C^{\mu\nu\rho\sigma}\sim f^{-2}(y)$ \cite{OrtPraPra11}, zeros of $f(y)$, which occur iff $\kappa K-8<0$, give rise to curvature singularities. The (singularity-free) $\Lambda<0$ solution with $\kappa K-8>0$ was already obtained in \cite{AyoCanZan16} and interpreted there as a traversable AdS wormhole with a NUT parameter $q$.

\section{Conclusions}

\label{sec_concl}

We have explored the generalized hedgehog ansatz and some extensions thereof in order to find exact solutions of the Einstein-nonlinear $SU(2)$ $\sigma$-model. We have considered four different subclasses of solutions giving rise to examples of physical interest in various contexts. The toroidal and spherical ans\"atze (sections~\ref{sec_1ai} and \ref{sec_3b}) contains static black holes (with toroidal and spherical horizons, respectively) along with their extremal limits. Extensions of these ans\"atze also lead to new nutty generalizations and to larger classes of Robinson-Trautman and Kundt metrics, which describe time-dependent solutions. The circular ansatz (section~\ref{sec_circular}) was used to describe two kinds of uniform black strings, with $\Lambda=0$ and $\Lambda<0$, respectively. Finally, the hyperspherical ansatz (section~\ref{sec_hypers}) was naturally employed in the context of Bianchi~IX cosmologies, where the pionic field plays the role of an anisotropic fluid. This also led to some special stationary solutions of the Bianchi~IX class. Along with new solutions,\footnote{To our knowledge, solutions which have not appeared previously in the literature include: \eqref{NUT_flat}, the general time-dependent ones of sections~\ref{subsec_RT_flat}, \ref{subsec_K_flat}, \ref{subsec_RT_spher} and \ref{subsec_K_spher}, \eqref{string1} (the latter is, however, implicitly contained in the construction of \cite{Giacominietal18}) and those of section~\ref{subsec_stationary_Bianchi} (except for the case $\kappa K-8>0$ \cite{AyoCanZan16}).} certain previously known ones (cf. the text for the corresponding references) have been rederived in a more systematic way. Let us observe that we have explored only some possible subclasses of the proposed ansatz, which means that a more systematic analysis might lead to further explicit solutions.

Until very recently it was considered as prohibitively difficult to find exact solutions of the self-gravitating nonlinear $\sigma$ (or Skyrme)-model. 
However, in the last years it has been shown that, by generalizing the hedgehog ansatz, the matter and Einstein field equations simplify dramatically \cite{CanSal13,CanMae13,AyoCanZan16,Astorinoetal18,Astorinoetal18_prd,Canforaetal19}. Our work provided further results in this direction. It is worth pointing out the hedgehog ansatz is also useful for finding exact solutions in Einstein Yang-Mills theory. For example, in \cite{CCGO} an explicit meron black hole solution for the self-gravitating $SU(2)$ Yang Mills theory has been found. This suggests that some of the techniques developed in this paper can also be applied in different contexts, such as the Skyrme and Yang-Mills theories coupled to gravity -- see, e.g., \cite{CanSal13,CanMae13,Astorinoetal18_prd} for a few cases where such a method indeed works. Exploring such extensions, however, will deserve a separate investigation.

\subsection*{Acknowledgements}

We are grateful to Marco Astorino and Fabrizio Canfora for stimulating discussions and to Malcolm MacCallum for helpful comments on \cite{MacSteSch70}. We also thank Adolfo Cisterna for useful remarks on a first draft of the manuscript. A.G. was supported by FODECYT project 1150246. M.O. was supported by research plan RVO: 67985840 and research grant GA\v CR 19-09659S. Part of the work of M.O. was carried out at Instituto de Ciencias F\'{\i}sicas y Matem\'aticas, Universidad Austral de Chile with the support of CONICYT PAI ATRACCI{\'O}N DE CAPITAL HUMANO AVANZADO DEL EXTRANJERO Folio 80150028.

\renewcommand{\thesection}{\Alph{section}}
\setcounter{section}{0}

\renewcommand{\theequation}{{\thesection}\arabic{equation}}

\section{Extended spherical ansatz: {\em test} fields in Kerr spacetime}
\setcounter{equation}{0}

\label{app_kerr}

The ansatz \eqref{metric3b}, with the additional restriction $g_{rG}=0=g_{tr}$, reduces to
\be
 ds^2=g_{tt}dt^2+2g_{tG}dtdG+g_{GG}dG^2+g_{rr}dr^2+g_{FF}dF^2 ,
\ee
where all the metric coefficients are functions only of $(r,F)$. If $g_{tt}<0$, this metric is stationary and axisymmetric\footnote{More rigorously, one should also impose regularity conditions at the axis, cf., e.g., \cite{Stephanibook}.} and clearly includes the Kerr(-Newman-(A)dS) spacetime in Boyer-Lindquist coordinates. However, it is easy to see that the field equation \eqref{D2F} is {\em not} satisfied when $a\neq0$ (not even in the flat case $m=0$) -- for the case $a=0$ see instead section~\ref{subsec_sphericalBH}.

Nevertheless, it turns out that an extended ansatz enables one to have the pionic field equations satisfied identically. Let us consider the Kerr metric in Kerr-Schild coordinates \cite{Kerr63}, i.e.,
\beqn
  & & ds^2=\Sigma(d\theta^2+\sin^2\theta d\phi^2)+(du+a\sin^2\theta d\phi)(2dr-du+a\sin^2\theta d\phi)+\frac{f}{\Sigma}(du+a\sin^2\theta d\phi)^2 , \label{kerr} \\
  & & \Sigma=r^2+a^2\cos^2\theta , \qquad f=2mr . \label{kerr_f}
\eeqn
Upon taking in \eqref{pions2} $\alpha=\pi/2$ and identifying the remaining pionic fields with the angular coordinates, i.e.,
\be
 (x^1,x^2,x^3,x^4)\equiv(\phi,\theta,u,r)=(G,F,u,r) ,
\ee
one can see that all pionic equations \eqref{fequ1}--\eqref{fequ3} are obeyed.\footnote{Note that this is an {\em extension} of the ansatz \eqref{metric3b} because here $g^{Gr}\neq0$ (as opposed to \eqref{gGF_gGr}) -- nevertheless, one still has $(\sqrt{-g}g^{Gr})_{,r}=0$ and thus $\square G=0$.} This is in fact true regardless of the choice of the function $f$ in \eqref{kerr} (hence, e.g., also in a Kerr-Newman background). Let us emphasize that these test field configurations are different from the ones recently constructed (in the Einstein-Skyrme theory) in \cite{Herdeiroetal18}, for example because of the different fall-off properties at infinity.

The next step would be to go beyond the test field approximation and take into account also the backreaction. Similarly as in the Kerr-Newman case \cite{Newmanetal65}, one would hope to preserve the general form of the line-element~\eqref{kerr} and only modify the scalar function $f$ in \eqref{kerr_f} in such a way that Einstein's equations are satisfied. However, this turns out to be inconsistent. It might thus be necessary to modify the form of the Kerr-Schild vector in \eqref{kerr}, or perhaps to go beyond the Kerr-Schild ansatz.

%
%
%
%

\end{document}